\documentclass[10pt]{article}

\usepackage{emulateapj}
\usepackage{psfig}
\usepackage{natbib}

\begin{document}

\newcommand{\J}{\mbox{${J1903+0327}$}}
\newcommand{\be}{\begin{eqnarray}}
\newcommand{\ee}{\end{eqnarray}}
\newcommand{\etal}{{\it{et al.}}}
\newcommand{\smass}{M_{\odot}}
\newcommand{\br}{{\bf r}}
\newcommand{\bV}{{{\bf v}}}
\newcommand{\Porb}{\mbox{${\rm P}_{\rm orb}$}}
\newcommand{\Msun}{\mbox{${\rm M}_\odot$}}
\newcommand{\Lsun}{\mbox{${\rm L}_\odot$}}
\newcommand{\Rsun}{\mbox{${\rm R}_\odot$}}

\renewcommand{\thefootnote}{\alph{footnote}}

\newcommand{\kms}{\mbox{${\rm km~s}^{-1}$}}
\newcommand{\msun}{\mbox{${\rm M}_\odot$}}
\def\apgt{\ {\raise-.5ex\hbox{$\buildrel>\over\sim$}}\ }
\def\aplt{\ {\raise-.5ex\hbox{$\buildrel<\over\sim$}}\ }



\def\simon#1{{\bf[#1 -- Simon]}}
\def\Simon#1{{\bf[#1 -- Simon]}}

\def\rev#1{{\bf #1}}

%
%

\title{The formation of the eccentric-orbit millisecond pulsar
  J1903+0327 and the origin of single millisecond pulsars}

%
%
\author{S.\, Portegies Zwart$^1$,
        E.P.J.\, van den Heuvel$^2$,
        J.\, van Leeuwen$^3$,
        and G.\, Nelemans$^4$ \\
$^1$Leiden Observatory, Leiden University,
           P.O. Box 9513, 2300 RA Leiden, The Netherlands \\
$^2$ Astronomical Institute `Anton Pannekoek', 
                Science Park 904
                1098 XH Amsterdam, The Netherlands \\
$^3$ Stichting ASTRON, PO Box 2, 
                7990 AA Dwingeloo, The Netherlands \\
$^4$ Dept. of Astrophysics, Radboud University Nijmegen, 
                Heyendaalseweg 135, NL-6525 AJ Nijmegen, 
                the Netherlands }
\date{}
%
%

\begin{abstract}
The millisecond pulsar J1903+0327 is accompanied by an ordinary
G-dwarf star in an unusually wide ($P_{\rm orb} \simeq 95.2$\,days)
and eccentric ($e \simeq 0.44$) orbit.  The standard model for
producing MSPs fails to explain the orbital characteristics of this
extraordinary binary, and alternative binary models are unable to
explain the observables.  We present a triple-star model for
producing MSPs in relatively wide eccentric binaries with a normal
(main-sequence) stellar companion. We start from a stable triple
system consisting of a Low-Mass X-ray Binary (LMXB) with an orbital
period of at least 1~day, accompanied by a G-dwarf in a wide and
possibly eccentric orbit.  Variations in the initial conditions
naturally provide a satisfactory explanation for the unexplained
triple component in the eclipsing soft X-ray transient 4U~2129+47 or
the cataclysmic variable EC 19314-5915.  The best explanation for \J\,
however, results from the expansion of the orbit of the LMXB,
driven by the mass transfer from the evolving donor star to its
neutron star companion, which causes the triple eventually to becomes
dynamically unstable.  Using numerical computations we show that,
depending on the precise system configuration at the moment the triple
becomes dynamically unstable, the ejection of each of the three
components is possible. If the donor star of the LMXB is ejected, a
system resembling \J\, will result. If the neutron star is ejected, a
single MSP results.  This model therefore also provides a
straightforward mechanism for forming single MSP in the Galactic disk.
We conclude that the Galaxy contains some 30--300 binaries with
characteristics similar to \J\, and about an order of magnitude fewer
single millisecond pulsars produced with the proposed triple scenario.
\end{abstract}


\section{Introduction}

The classic channel for the formation of a millisecond pulsar (MSP)
requires a close binary with an extreme mass ratio ($\aplt
2/10$). This binary survives a common-envelope evolution and the
subsequent supernova explosion of the primary star.  The neutron star,
formed in the supernova, can subsequently be spun up to a millisecond
pulsar (MSP) in a phase of mass transfer from the $\aplt 2$\,\Msun\,
Roche-lobe filling companion star, which in the process is slowly
stripped from its envelope. During this phase the binary is visible as
a low-mass x-ray binary (LMXB), eventually resulting in a MSP that is
accompanied by a low-mass white dwarf in a relatively wide, almost
circular orbit ($e \aplt 10^{-3}$) \citep{1991PhR...203....1B}. In the
Galaxy 50 such systems are known, while about 15 MSPs in the Galaxy
have no companion at all \citep{LorimerLivingReview2008}.

The formation of the recently observed binary millisecond pulsar
\J\ cannot be reconciled with the above scenario. It's characteristics
are too different: the companion star is a G-dwarf instead of a white
dwarf, and the orbit is highly eccentric, $e \simeq 0.44$ instead of
the expected $\aplt 10^{-3}$ \citep{1998ApJ...505..315C}.  In
addition, the average mass of the companion of known Galactic MSPs
with pulse period $<10$\,ms is $0.22\pm 0.17$\,\Msun, whereas for \J\,
the companion mass is 1.03\,\msun\, \cite{2010arXiv1011.5809F}; and
binary MSPs with a companion mass $>0.6$\,\Msun\, tend to have a long
pulse period $\langle P \rangle = 62 \pm 76$\,ms and short orbital
periods ($P_{\rm orb} = 5.2\pm 4.8$\,days), whereas \J\, has an
extremely short pulse period of 2.15\,ms and an extraordinary long
orbital period of 95\,days.  These discrepancies with respect to the
expected outcome of the standard scenario for producing a millisecond
pulsar in a binary requires an exotic solution.

We propose that \J\, was born as a rather ordinary triple star of
which the inner binary is the progenitor of a LMXB, and with an outer
(tertiary) star that initially is less massive than the secondary so
that the inner secondary evolved first. After a common-envelope phase
and a supernova explosion, mass transfer in the inner binary leads to
expansion of its orbit. Depending on the orbit of the outer star a
dynamical instability ensues in which one of the three stars is
ejected. Such an evolution results in a MSP binary with an outer
companion in either a wide orbit (if the instability is avoided), or a MSP
binary if the outer companion is ejected, or a normal binary plus a
single MSP if the neutron star is ejected, or if the inner secondary
is ejected, a MSP with a low-mass companion in an eccentric orbit,
like \J. 

The scenario sounds exotic, and it is, but in
\S\,\ref{Sect:InitialConditions} and \S\,\ref{Sect:HowManyJsInGalaxy}
we estimate the available parameter space and calculate that this
model leads to an acceptable birthrate.  (We notice that in a recent
paper \cite{2010arXiv1011.5809F} independently also suggest this model
one of possible triple star models for \J.)  We will now dwell on the
details of our scenario in \S\,\ref{Sect:ProposedModel} and quantify
the results by performing simulations of triple star systems in
\S\,\ref{Sect:Simulations}.  We discuss the shortcomings
of earlier proposed scenarios for the formation of \J\, in
\S\,\ref{Sect:Discussion} and summarize our conclusions in
\S\,\ref{Sect:Conclusions}

\section{The triple scenario for forming \J}\label{Sect:ProposedModel}

We propose that the binary \J\, was born as a rather ordinary triple
star. We envision a configuration at birth where a relatively massive
9-12\,\Msun\, primary star was orbited by a secondary star of $\sim
0.8$ to 2.0\,\msun\, in a relatively close ($\sim 200$\,\Rsun) orbit,
and a tertiary star which is less massive than the secondary and has a
rather wide $\apgt 560$\,\Rsun\, orbit around the inner binary.
Adopting a tertiary mass that exceeds the secondary would considerably
change the outcome of the evolution, because in that case it is the
outer star that ascends the giant branch and possibly fills its Roche
lobe before the inner binary has turned into a LMXB. The evolution of
such triples has not yet been subject to any study
\citep{1996epbs.conf..345E,2006epbm.book.....E}.

When the massive primary of the initial triple ascends the red-giant
branch a common-envelope ensues
\citep{1984ApJ...277..355W,2000ARA&A..38..113T,2010ApJ...719L..28D,2010ApJ...717..724G},
in which the inner secondary and the degenerate core of the giant
spiral in towards each other. A few Myr later, the stellar core
explodes as a supernova, forming a neutron star with a low-mass
companion in an eccentric orbit with a period of a few to several tens
of days \citep{2010ApJ...719L..28D}. The outer orbit is likely
influenced both by the mass shell ejected in the common envelope, as
well as by the supernova explosion. The mass shell containing 6 to
10\,\Msun, may slow down the outer companion star, which may cause its
orbit to shrink, but the removal of mass from the system may also
widen it. The supernova explosion causes the inner binary to receive a
velocity kick \citep{1961BAN....15..265B,1987ApJ...321..780D} which
changes the orbits and may disrupt the triple.  The kick results from
two subsequent effects; the mass loss in the explosion and the
intrinsic velocity kick imparted to the newly formed neutron star (see
\S\,\ref{Sect:TripleSurvivors} for details).  In the case of an
electron-capture supernova the latter effect is expected to result in
a relatively low-velocity kick $<50$\,km/s
\citep{2005MNRAS.363L..71D}, which generally sufficices to keep the
outer orbit bound by compensating the effect of the mass loss in the
supernova. In the surviving systems mass transfer will ensue in the
inner binary after several Gyr. This rapidly circularizes the inner
orbit by tidal forces and turns the system in a Low Mass X-ray Binary
(LMXB) resembling the bright Galactic bulge LMXBs
\citep{1983ApJ...270..678W,1983ApJ...270..694T}.  An example of such a
triple is 4U~2129+47 (V1727 Cyg), in which a 5.24 hour LMXB is
accompanied by a spectral type F-dwarf in an eccentric orbit of about
175\,days
\citep{1989ApJ...341L..75G,2008A&A...485..773B,2009ApJ...706.1069L}.
For our model we require that, contrary to 4U~2129+47, at the moment
of RLOF the LMXB has a period of at least $\sim 1$\,day and will
evolve to longer periods, rather than shrink due to the emission of
gravitational waves
\citep{1988A&A...191...57P,1989A&A...208...52P}. In the case of
4U~2129+47, the inner orbit will not expand due to the mass transfer
and therefore will not perturb the outer orbit.

Mass transfer in the inner binary causes the accreting neutron star to
be spun up to a millisecond spin period
\citep{1982Natur.300..728A,1982CSci...51.1096R}. After several tens of
millions of years, the expansion of the inner orbit leads to a
dynamically unstable situation with respect to the orbit of the outer
star. Depending on the precise configuration at the moment when the
system becomes dynamically unstable, either the donor of the LMXB, or
the outer tertiary star or the neutron star can be ejected from the
unstable triple. In the latter case a single MSP results, in the
first-mentioned case a wide eccentric millisecond binary pulsar with a
G-dwarf companion is produced, resembling \J.

\section{Constraining the birth conditions for \J}\label{Sect:InitialConditions}

We investigate the conditions under which the proposed triple scenario
produces a system similar to \J. With the currently observed orbital
parameters of \J\, ($a \simeq 123.2$\,\Rsun, $e \sim 0.44$,
$m_G\simeq1.03$\,\Msun\, and $M_{\rm MSP}\simeq1.67$\,\Msun\,
\citep{2010arXiv1011.5809F}) the binding energy of the binary is about
$E_b \simeq 2.65\times 10^{46}$\,erg.  This poses a minimum to the
binding energy of the triple just before it became dynamically
unstable, and allows us to calculate the orbital separation of both
the inner and the outer orbit at the moment that the dynamical
instability sets in.  In practice the binding energy of the stable
triple will be higher than $E_b$ by the ratio of the masses of the
ejected star with respect to the triple, or $\sim 20$\%
\citep{1975MNRAS.173..729H,2004MNRAS.352....1F}, because the escaping
star carries off some fraction of the binding energy to infinity.

The requirement that the binding energy of the triple must exceed
$E_b$ allows us to calculate the orbital separation of both the inner
and the outer orbit at the moment that the dynamical instability sets
in. The criterion for dynamical stability sensitively depends on the
separation and the eccentricity of the inner and outer orbits, and on
the masses of the three stars \citep{2001MNRAS.321..398M}.  However,
for the observed system \J\, only the masses of two of the stars (the
MSP and its current G-dwarf companion) are known, as the initial
secondary star was ejected from the binary.

The initial donor in the LMXB must have been more massive than the
current MSP companion in order to evolve first, but not so massive
that mass transfer in the LMXB would be unstable; we therefore adopt a
mass of the original close companion of the neutron star of 1.0 --
2.0\,\Msun\, (In principle a secondary mass as low as 0.8\,\msun\,
would suffice to warrant the inner binary to evolve into a LMXB, but
in that case the tertiary star should be $< 0.8$\,\msun, which is
smaller than than the observed 1.03\,\msun). For the neutron star we
adopt a mass of 1.30\,\Msun\, before it starts accreting mass
\citep{2010ApJ...719..722S}, and in the same range of masses, but less
massive than the inner secondary star for the outer companion. In the
specific case of \J\, the observed star is an 1.03\,\msun\, G-dwarf,
which in our scenario would be identical to the initial tertiary
star. The best match to the orbital characteristics are then obtained
when we adopt the initial secondary mass to be $\sim 1.1$\,\msun.

During the LMXB phase mass is transferred from the inner companion
(donor) to the neutron star.  The time averaged mass-transfer rate can
be estimated from the empirical relation based on the initial orbital
period of the LMXB \citep{1994inbi.conf.....S}. In our model we assume
that the neutron star accretes at most at the Eddington limit; for
which we adopted $\dot{m} = 1.5 \times 10^{-8}$\,\Msun/yr.  The
left-over mass that is provided by the donor but that is not accreted
by the neutron star is assumed to leave the inner binary with the
specific orbital angular momentum of the neutron star
\citep{1995A&A...296..691P}.  The currently observed mass of the
neutron star is 1.67\,\Msun, which indicates that in our model it must
have accreted about 0.37\,\Msun, and that the donor star must have
lost at least that same amount of mass. Mass transfer in such a binary
system causes the orbit to expand quite dramatically, in particular if
the mass transfer was not conservative.

\begin{figure}
  \includegraphics[width=0.45\textwidth]{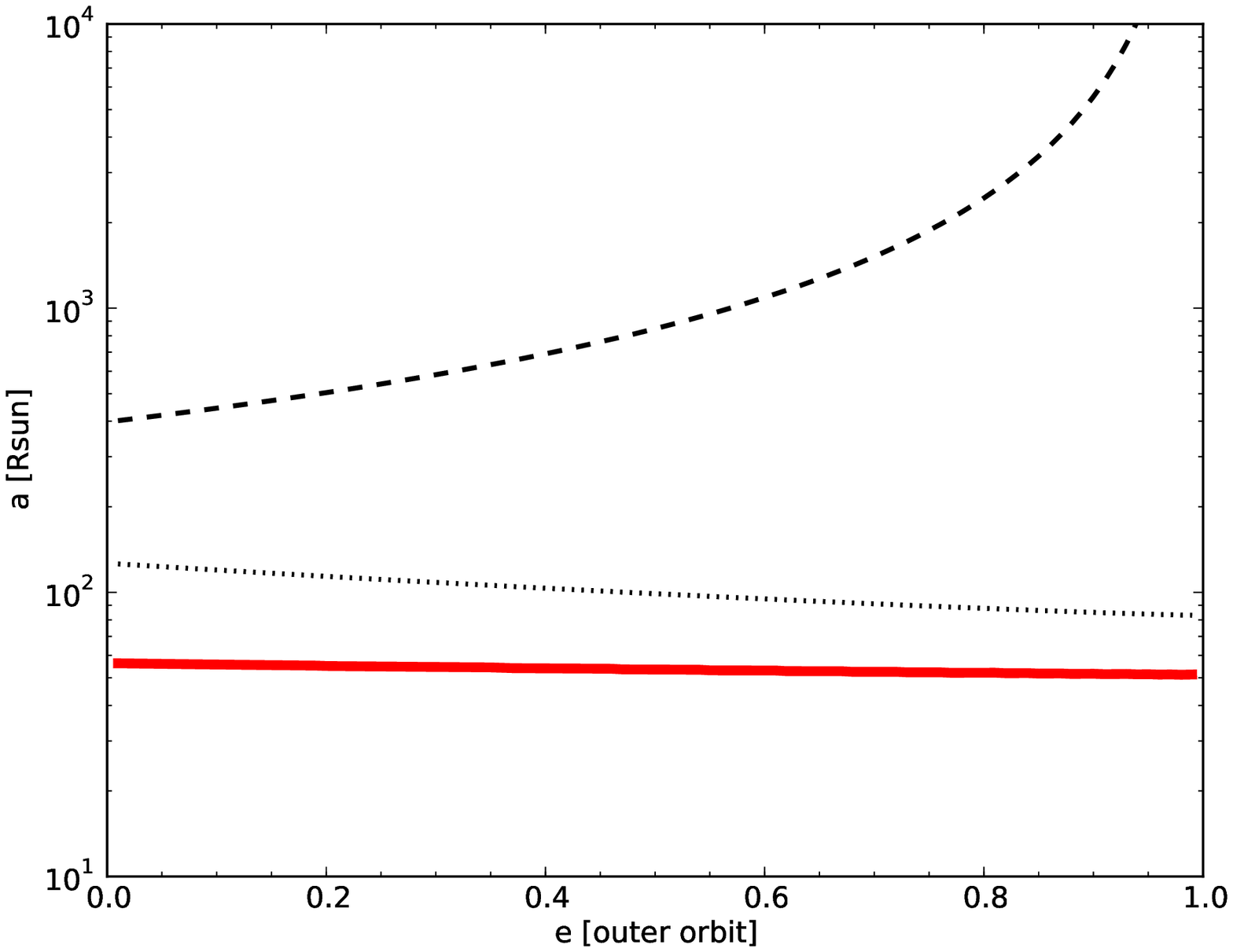}
\caption[]{A triple with a 1.3\,\Msun\, neutron star and 1.1\,\Msun\,
  G-dwarf in a circular inner orbit (thick solid curve [red]) and a
  1.0\,\Msun\, G-dwarf in an eccentric outer orbit (dashes) is stable
  and has the same binding energy as \J.  These curves are calculated
  using the iterative procedure described in
  \S\,\ref{Sect:InitialConditions}. The orbit of the LMXB expands as a
  consequence of the mass that is transferred from the initial
  secondary donor star to the neutron star. According to the analytic
  expression provided by \cite[][their Eq.\,90]{2001MNRAS.321..398M},
  the triple becomes dynamically unstable as soon as the orbital
  separation of the LMXB reaches the dotted curve. At this moment the
  triple dissolves and one of the three stars is ejected, leaving
  behind a binary with an orbital separation roughly somewhere between
  the dotted and the dashed (black) curve (see
  Fig.\,\ref{Fig:Fig_ae_distribution}).}
\label{Fig:BindingEnergyAnalysis}
\end{figure}

We can now determine the separation of the inner and the outer orbit
for the initial triple (before the LMXB-phase) by an iterative
procedure under the constraints that the total binding energy of the
triple at the moment it became dynamically unstable is known, and by
taking the effect of the mass-transfer process during the LMXB phase
into account. Here we assumed that the mass lost from the LMXB leaves
the triple in the form of a stellar wind.  The resulting orbital
constraints of this iterative procedure are presented in
Fig.\,\ref{Fig:BindingEnergyAnalysis} as the solid and dashed curves;
these give the most likely initial conditions for forming \J.

\placefigure{Fig:BindingEnergyAnalysis}

With the semi-analytic procedure just described we still cannot
predict the consequence of the dynamical instability, in particular
because the final stage of the triple is highly dynamic and the
orbital parameters do not allow us to predict the identity of the
ejected star from first principles. However, one can imagine that it
is quite likely that the less massive donor ($<0.73$\,\Msun\, when we
adopt an initial secondary mass of 1.1\,\msun) is ejected, rather than
the more massive MSP ($\sim 1.67$\,\Msun) or the outer star ($\sim
1.03$\,\Msun).  At the moment when the dynamical instability sets in the
donor still has part of its Hydrogen envelope, which explains why its
mass exceeds that of the degenerate Helium core.

\section{Simulating mass transfer in a triple system}\label{Sect:Simulations}

We quantify our proposed formation channel for \J\, 
by performing extensive computer simulations of triple systems, starting at the onset of the LMXB phase.  The
calculations ware performed using the Astrophysics Multipurpose
Software Environment (AMUSE)\footnote{see {\tt
    http://www.amusecode.org}}\citep{2009NewA...14..369P}.

In the AMUSE environment we resolve the dynamical evolution of the
triple-star by a specialized numerical orbit integration, which is
written in FORTRAN77.  The mass transfer in the inner binary is
implemented in python. The outer star is treated as a point-mass and
was not evolved during the simulation.  The coupling between the
numerical orbit integration of the three stars and the stellar
evolution calculations is realized using AMUSE
\citep{2009NewA...14..369P}.  The most important role of AMUSE is
converting the units and to realize the communication between the two
codes. The former is done with a specialized unit conversion module
and for the latter we spawned the different processes using
the Message Passing Interface (MPI) to transfer the required data in
the proper units and in discrete instances between the two codes.

\subsection{Numerical Method} \label{Sect:NumericalMethod}

The orbits of the triple stars were integrated using a regularized
version of the Burlish-Stoer integrator
\citep{1974CeMec..10..185A,1974CeMec..10..516A}, keeping the numerical
error at machine precision, and allowing a maximum relative energy
error of {\large O}($10^{-14}$) per integration of the outer orbit.

During the dynamical evolution we resolved the mass transfer and
consequential change in orbital paramters of the LMXB that is orbited
by the outer star.

The rate of mass-transfer in the inner binary is calculated from the
empirical relation based on the orbital period at the onset of RLOF in
the LMXB: $\langle \dot{m} \rangle = 6 \times 10^{-10} (\Porb ({\rm
  initial})/1 {\rm day})$\,\Msun/yr \citep{1994inbi.conf.....S}.

The neutron star was allowed to accrete at most at the Eddington rate
and any surplus mass is assumed to leave the binary with the
angular momentum of the accreting neutron star, but lost adiabatically
from the triple. 

We perform the mass transfer in the inner binary every time the outer
orbit has had 10 revolutions, after which the numerical orbit
integration was updated using the newly calculated orbital parameters
which resulted from carrying out the mass transfer.  We varied the
interval between which mass transfer in the inner binary was conducted
between every 1 to 1000 outer orbits, but this choice did not
significantly affect the results.  (In \S\,\ref{Sect:JlikeSystems} we
present the results of a series of simulations where we decoupled the
mass-transfer process from the gravitational evolution after the
triple has become dynamically unstable.)

\subsection{Results of the simulations} \label{Sect:SimulationResults}

We initialized $10^3$ binaries and calculated the evolution for each
up to an age of at most 10\,Gyr or until the mass of the donor star
drops to the mass of the degenerate helium core for a population-II
star \citep[$\aplt 0.4$\,\msun,][]{1999A&A...350..928T}.

Each triple was initialized by randomly selecting the eccentricity of
the outer orbit from the thermal distribution and the initial
separations of the inner and outer orbits for each triple-evolution
calculation are selected using the iterative procedure described in
\S\,\ref{Sect:InitialConditions} (see the thick solid and dashed
curves in Fig.\,\ref{Fig:BindingEnergyAnalysis}).  For each simulated
binary we randomly selected the inclination of the inner orbit with
respect to the outer orbit, the longitude of the ascending node, the
argument of periastron and the phases of the two orbits.  The
preference in the orbital elements introduced by the supernova
explosion may affect the rate of ejected MSPs relative to those that
stay in a binary, but we ignore that complication.  We found no
significant correlations between the final outcome of the simulations
and longitude of the ascending node, the argument of periastron or the
phases of the two orbits. The anisotropic velocity caused by the mass
loss in the supernova of the inner binary therefore is not expected to
have a significant effect on the surviveability of the triple.

During the evolution of the triples, small eccentricities ($\aplt
0.1$) are commonly induced in the inner orbit, in particular when the
triple approaches the regime where it becomes dynamically unstable.
We stop a simulation when the eccentricity of the inner (LMXB) orbit
exceeds 0.3 for more than $10^5$ years. Such high eccentricities can
be induced shortly by a strong interaction with the outer star, which
typically results in the break-up of the triple, or by a long term
secular perturbations of the inner orbit by the outer star
\citep{1979IAUS...81..231K}.  In the latter case orbital variations
result naturally from the secular evolution of the triple and do not
directly lead to a dynamical instability. We still decided to
terminate such simulations because it becomes hard to follow the
mass-transfer process within the inner binary. It would require
extensive hydro-dynamical simulations to study the consequences of a
Kozai resonances in a triple with a Roche-lobe filling inner
binary. About 10\% of our simulations ware stopped as a consequence of
this effect.  Mass transfer in eccentric orbits is generally ill
understood, although courageous attempts are underway to tackle this
problem
\citep{2009ApJ...702.1387S,2010ApJ...724..546S,2010arXiv1011.2204L,2010arXiv1011.2211L}.

The majority (917) of the binaries become dynamically unstable long
before the other stopping criteria apply, in which cases we continue
to resolve the dynamics by integrating the equations of motion and
resolving the internal mass transfer until one of the stars
escapes. The orbital separation of the LMXB at which the triple is
expected to become unstable is indicated by the dotted curve in
Fig.\,\ref{Fig:BindingEnergyAnalysis}.

\begin{figure}
  \includegraphics[width=0.45\textwidth]{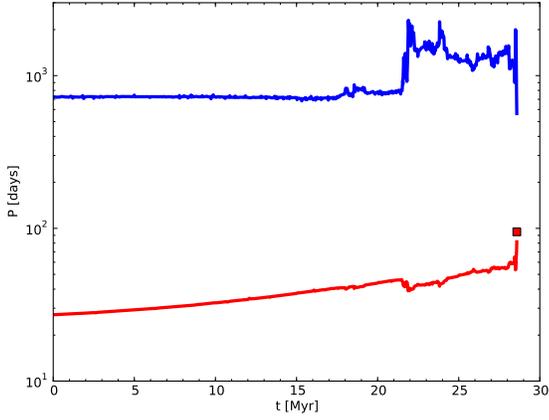}
\caption[]{Evolution of the orbital period of a rather typical case
  which evolves to a system very similar to \J.  The square (red)
  represents the final orbit of this simulation, which in this
  particular case resulted in a binary consisting of a 1.73\,\Msun\,
  MSP and a 1.0\,\Msun\, companion in a $\sim 100$\,day orbit with
  $e\simeq0.43$.  The bottom line (red) indicates, as a function of
  time the evolution of the inner LMXB, whereas the top line (blue)
  represents the evolution of the outer orbit.  }
\label{Fig:ExampleEvolution}
\end{figure}

An illustrative example of the evolution of the period of the inner
and outer orbits is presented in Fig.\,\ref{Fig:ExampleEvolution}.
The evolution of the inner LMXB also drives the expansion of the outer
orbit, in particular by the dynamical coupling between both orbits and
in a lesser extend by the mass lost from the inner LMXB in those cases
that mass transfer proceeds non-conservatively.  This relative
softening causes the final MSP binaries to be somewhat wider than
observed in \J, and as a consequence the triple remains stable for
somewhat longer than expected based on our analytic energy balance
(see \S\,\ref{Sect:InitialConditions}), and eventually results in the
MSP to be somewhat more massive (by about 0.2\,\Msun) than observed in
\J. We can compensate for this by adopting a slightly ($\sim 20$\,\%)
smaller outer orbital separation at the onset of the LMXB phase.

\placefigure{Fig:ExampleEvolution}

\begin{figure}
  \includegraphics[width=0.45\textwidth]{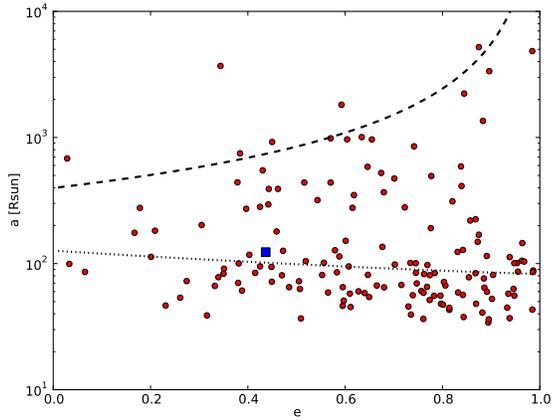}
\caption[]{The eccentricity ($e$) and semi-major axis ($a$) of the MSP
  binaries with G-dwarf companions that resulted our from simulations
  (dots [red], see \S\,\ref{Sect:InitialConditions}).  These binaries
  are initialized to mimic \J, and which are described in
  \S\,\ref{Sect:InitialConditions} and represented by the thick solid
  and dashed curves in Fig.\,\ref{Fig:BindingEnergyAnalysis}, and
  which are reproduced here to guide the eye regarding the most likely
  range of final binary parameters.  The square (blue) indicates the
  current orbital separation and eccentricity of \J.  Note that the
  final eccentricity of the surviving binary is unrelated to the
  eccentricity of the initial outer orbit (black dashed curve).  }
\label{Fig:Fig_ae_distribution}
\end{figure}

The resulting parameters of the MSP with G-star binary are presented
in Fig.\,\ref{Fig:Fig_ae_distribution}, and straddle the observed
orbital separation and eccentricity of \J.  The phase of mass transfer
for these binaries lasted for about $23.4\pm9.3$\,Myr, after which the
donor was ejected in about one-fifth $\sim 20$\% of the cases. The
resulting binaries, for those with $a<10^4$\,\Rsun, had an average
orbital separation of $175 \pm 145$\,\Rsun, and an eccentricity of
$0.65\pm0.19$.  During the LMXB phase, the neutron stars were able to
grow from 1.30\,\msun\, to $1.64 \pm 0.12$\,\msun. The mean mass of the
ejected donor was $0.72\pm0.13$\,\msun. 

Our success in reproducing the observed parameters of the binary MSP
\J\, demonstrates that its progenitor may well have been born as a
triple star.  However, the here described pin-pointed search of
parameter space makes it impossible to derive the birthrate for \J,
which we will calculate in the next \S.


\placefigure{Fig:Fig_ae_distribution}

\section{How many \J-like systems are there in the Galaxy?}\label{Sect:HowManyJsInGalaxy}

We calculate the formation rate of \J-like systems by determining
their birth rate with respect to that of ordinary LMXBs. The reason
for this approach is the great uncertainty in the number of LMXB
progenitors because binaries with such extreme mass ratios are
observationally unknown.

This ratio depends on several factors
\begin{enumerate}
\item The number of triple systems with suitable parameters.
\item The fraction of triples that ensues and survive the
  common-envelope phase and the supernova explosion in which the
  neutron stars is formed
\item The fraction of those triples that lead to \J-like systems
  (instead of single MSPs, classic MSP binaries or MSP binaries with a
  triple companion)
\end{enumerate}
We will discuss each these factors below, in \S\,
\ref{Sect:NumberOfTriples}, \S\,\ref{Sect:TripleSurvivors} and \S\,
\ref{Sect:JlikeSystems}, respectively.

\subsection{The number of suitable triple stars}\label{Sect:NumberOfTriples}

According to the Hipparcos database the ratio of hierarchical higher
order multiple stellar systems (404) to binaries (1438) is $404/1842
\sim 0.22$ and the outer star have a rather flat mass distribution
\citep{2008MNRAS.389..869E,2010yCat..73890869E}. We assume that for
(inner) binaries that are progenitors to LMXBs the same ratio holds,
even though there are none in this catalogue.  We further require that
the mass of the outer stars is about ten times lower than that of the
inner binary. We then find a ratio of suitable triples to LMXB
progenitor binaries of a few per cent.  We present in
\S\,\ref{Sect:MissingLinks} observational evidence that indeed such
triples exist.

\subsection{The fraction of triples that survive the supernova explosion} \label{Sect:TripleSurvivors}

We will not dwell on the details of the common envelope evolution, but
assume that it leads to the spiral-in of the inner two components
without much affecting the outer star. The consequences of the common
envelope do not qualitatively affect our result, but have a strong
effect on the derived birth rate (see \S\,\ref{Sect:Birthrate}). In
\S\,\ref{Sect:MissingLinks} we discuss that the orbit of the tertiary star
that orbits the LMXB 4U~2129+47 may have experienced a considerable
reduction during the common-envelope, but we consider this
insufficient evidence to draw general conclusions regarding the effect
of the common envelope on the outer orbit.

The effect of the supernova explosion in the inner binary on the
orbital parameters of the triple can be calculated relatively
straight-forward, and has a profound effect on the survivability of
the triple because the weakly bound outer orbit is easily disrupted.

\begin{figure}
$a$)~\includegraphics[width=0.45\textwidth]{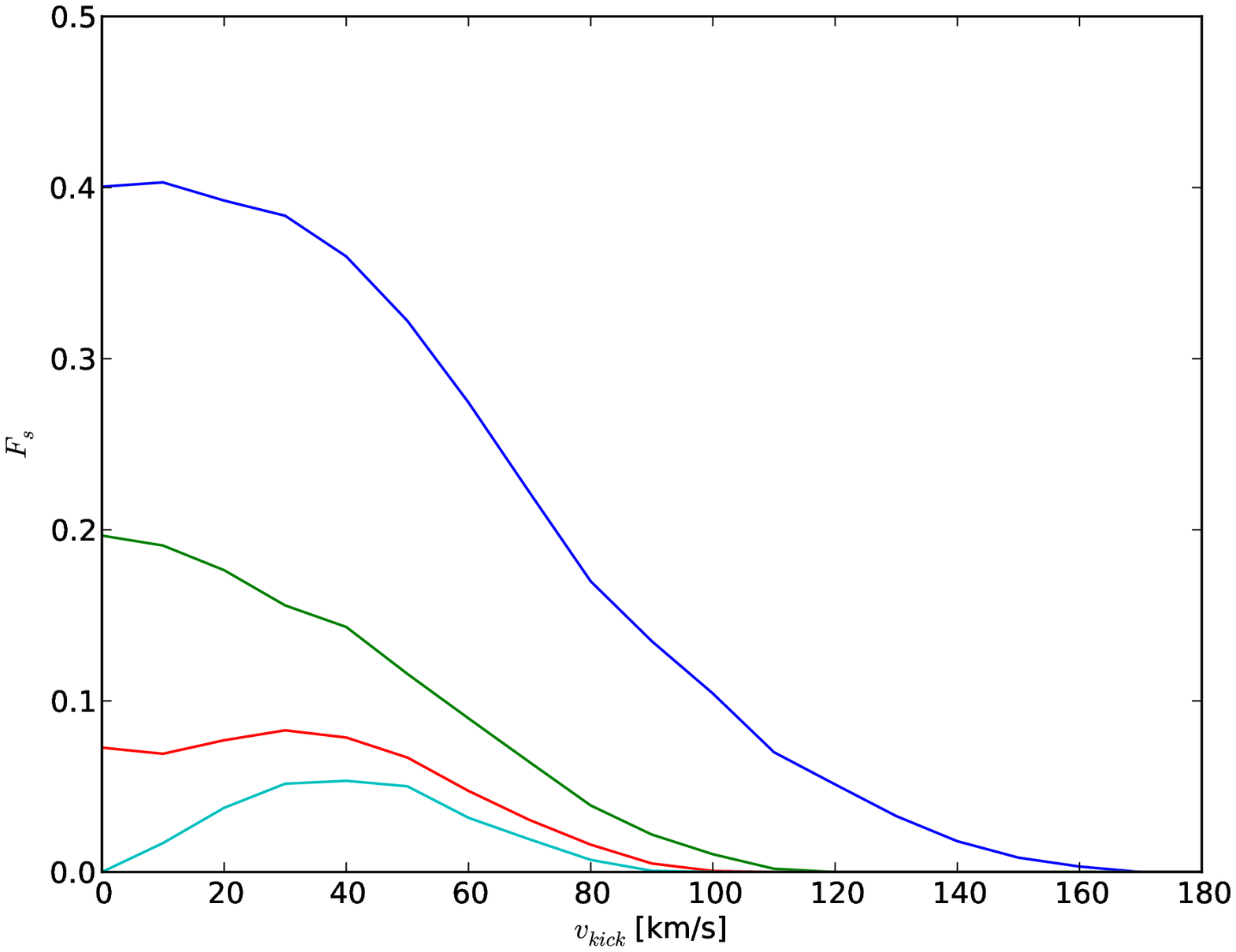}
$b$)~\includegraphics[width=0.45\textwidth]{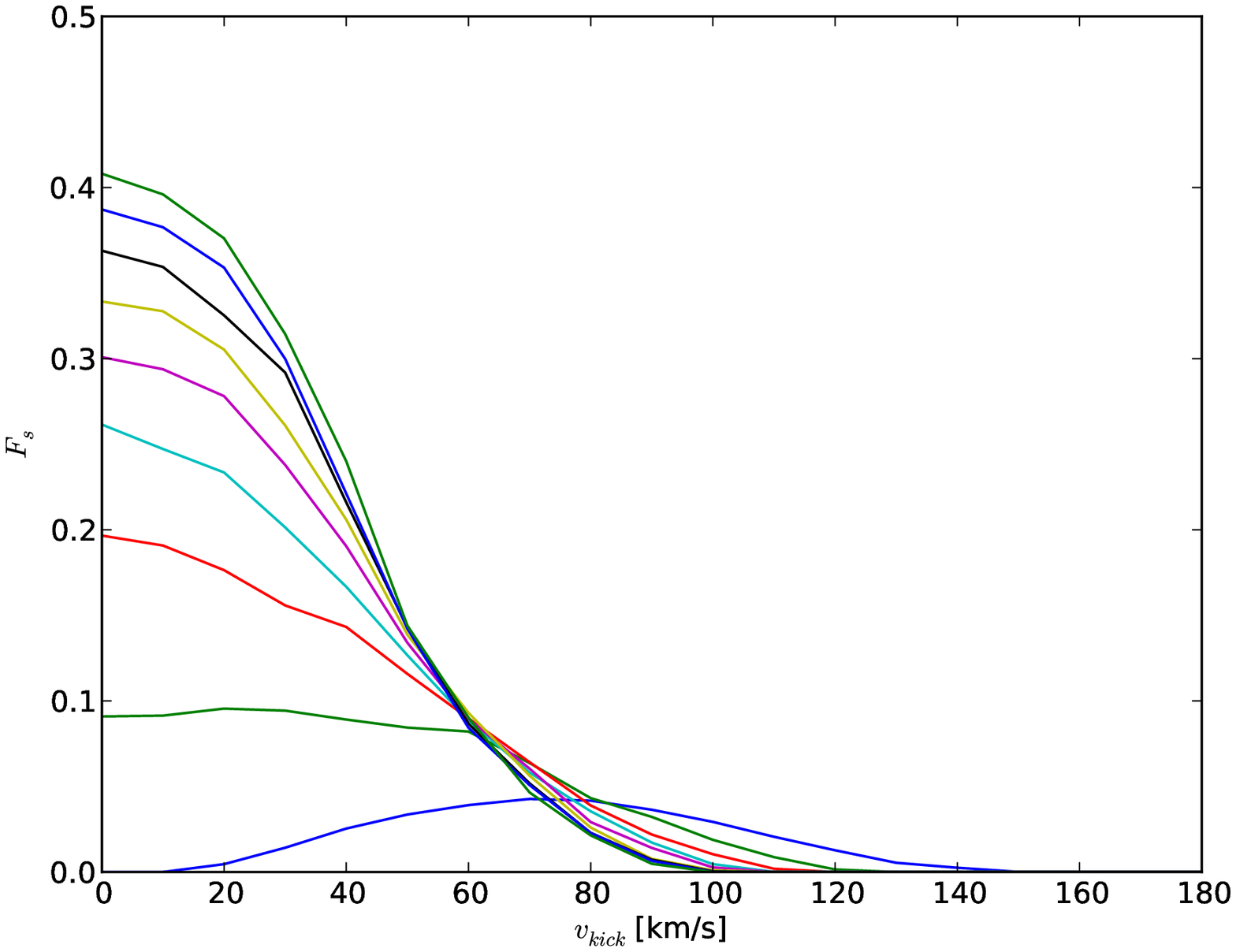}
\caption[]{Survival probability for a triple in which a 2.7\,\msun\,
  star explodes in a supernova explosion to a 1.3\,\msun\, neutron
  star. The inner companion star is 1.1\,\msun\, and the outer star is
  1.03\,\msun.  The inner orbit is circular and with a semi-major axis
  of 30\,\Rsun\, (panel $a$) and a circular outer orbit is
  1000\,\Rsun, 3000, ..., 9000\,\Rsun\, (top to bottom).  Panel $b$
  gives the supernova survival fraction for an outer separation of
  3000\,\Rsun\, and an inner orbit of 10, 20, ..., 90\,\Rsun\, (bottom
  to top curve). Note that we adopted here high mass-loss in the
  supernova contrary to an electron-capture supernova, which
  suppresses the survival rate.}
\label{Fig:SupernovaSurvival}
\end{figure}

The mass loss in the supernova explosion causes the inner binary to be
ejected \citep{1961BAN....15..265B,1961BAN....15..291B}.  In order to
keep the triple bound a small asymmetric velocity kick imparted to the
newly formed neutron star is required
\citep{1983ApJ...267..322H,1995MNRAS.274..461B,1998A&A...330.1047T}.
In Fig.\,\ref{Fig:SupernovaSurvival} we show the survival probability
of several combinations of inner and outer orbits as function of the
asymmetric kick magnitude. In these calculations the effect of the
Blaauw-Boersma kick was self-consistently taken into account to calculate
the survival probability.  This fraction ranges from $\sim 40$\% for
small kicks to zero for kicks significantly above $100\,\kms$.

\placefigure{Fig:SupernovaSurvival}

To study the future evolution of the triple through the LMXB phase,
we need to derive the orbital parameters as a function of the most
probable kick velocity and the amount of mass lost in the supernova.
We consider several combinations of both, and the distinction between
them can be made by the mass of the initial primary star and the
moment it filled its Roche-lobe.

In \S\,\ref{Sect:InitialConditions} we discussed the most likely range
of parameters at birth, and we adopt these same numbers here to study
the surviveability of the triple.  Here we make a distinction between
two types of supernovae; in one case we adopt that the inner primary at
the moment of the supernova is a 1.9\,\msun\ Helium star with a
degenerate ONeMg core that is the left-over of the initial
9--12\,\Msun\, primary star. If the initial primary star was born
somewhat more massive, 10--13\,\Msun\, and was stripped from its
Hydrogen envelope at a later stage of its evolution it's Helium core
may have grown to $\sim 2.7$\,\msun.  The ONeMg star that experiences
an electron-capture supernova loses $\sim 0.6$\,\msun\, and if the
exploding star has a more massive Helium core the mass loss is $\sim
1.4$\,\msun. In both cases we adopt a neutron star mass of
1.30\,\msun.

Apart from the smaller mass lost in the explosion of the ONeMg core,
the neutron star is also expected to receive a smaller velocity kick
upon birth. For these electron capture supernovae we adopted a
Gaussian distribution for the velocity kick with a dispersion of
20\,km/s \citep{2004ApJ...612.1044P,2004PhRvL..92a1103S} in a random
direction.  From studies of single radio pulsars in the solar
neighborhood several kick velocity distributions have been constructed
\citep{1994Natur.369..127L,1997MNRAS.291..569H,1998ApJ...505..315C,2002ApJ...568..289A},
most of them with considerbly higher velocity than for the ONeMg
supernovae.  These latter kick velocity distributions are expected to
be more suitable for neutron stars formed from an isolated star or a
non-interacting binary, in which case the neutron star is formen by
the collapse of an iron core
\citep{2004ESASP.552..185V,2004ApJ...612.1044P}.  These kicks range
from single (zero-centered) Gaussian distributions with a velocity
dispersion of 265\,km/s \cite{2005MNRAS.360..974H} to more complicated
distributions like the proposed two Gaussians with dispersions of
175\,km/s and 700\,km/s and relative probabilty of 0.86 and 0.14,
respectively \citep{1998ApJ...505..315C}.

We study the probability that a triple survives the supernova by means
of Monte-Carlo simulations.  The simulated triples
had the following characteristics: The inner binary consists of a
1.9\,\msun\ Helium star with a degenerate ONeMg core or a $\sim
2.7$\,\msun\, Helium star, and a 0.8 -- 2.0\,\msun\, secondary. The
latter was selected randomly with equal probability within the
interval.  The orbital separation of the circular inner binary was
taken flat in $\log$\, between 1\,\Rsun\, and 100\,\Rsun.  We adopted
these parameters from the population synthesis calculation of LMXBs
\citep{2003MNRAS.343..949W}, in particular using the results of their
models KM25 to KM100 of \citep{2003MNRAS.343..949W}.

The mass of the tertiary star was selected to be less massive than the
secondary, but not smaller than 0.8\,\msun\, (in theory there is no
lower limit for the mass of the tertiary star). The outer orbital
separation was chosen with a probability distribution flat in $\log$
with a maximum of $10^4$\,\Rsun.  The minimum separation was chosen to
be consistent with a dynamically stable initial triple (adopting a
10\,\msun\, primary and an inner orbital separation of 200\,\Rsun) and
the earlier selected secondary and tertiary masses. The latter two
stars ware assumed not to accrete any material throughout the common
envelope and supernova explosion. The eccentricity of the outer orbit
before the supernova was selected at random from the thermal
distribution between a circular orbit and a maximum which was chosen
such that the triple is dynamical stable. The selection of the minimum
orbital separation of the outer star and it's eccentricity therewith
becomes an iterative procedure. The other orbital elements are selected
randomly, as we described in \S\,\ref{Sect:SimulationResults}.

For each system we calculate the effect of the combined Blaauw-Boersma
and intrinsic velocity kick on the inner and the outer orbit, the
latter kick was assumed to be isotropic. In our simulations we varied the
mass loss in the supernova and the velocity distribution of the
asymmetric kick. For electron-capture supernovae
\citep{2004ApJ...612.1044P} the fraction of surviving triples is
1/3. For higher kick velocities we adopted that the exploding star was
2.7\,\msun\, and as a consequence the fraction of survivors drops to
1/25 \citep{2002ApJ...568..289A}, 1/28 \citep{2005MNRAS.360..974H} and
1/50 \citep{1998ApJ...505..315C} where the quoted literature refers to
the adopted kick velocity distribution.  Note here that the smaller
amount of mass lost in the electron-capture supernova explosions helps
considerably in preserving more triples, as opposed to the more
violent kicks.

We conclude that for every 3 inner binaries that survive the
electron-capture supernova the outer tertiary star remains in orbit
around the inner binary, but that this fraction may drop considerably
(to 1/50) when more mass is ejected in the supernova shell and the
kick velocity is higher.  Varying the amount of mass lost in the
supernova explosion has a profound effect on the survivability of the
triple, in particular since the Blaauw-Boersma kick imparted on the
inner binary is proportional to this mass loss, and to the relative
orbital velocity of the inner binary.  It is interesting to note that
the vast majority of the triples that survive the supernova explosion
are dynamically stable, but their orbital eccentricity tend to be
considerably higher than according to the thermal distribution.

\subsection{The fraction of surviving triples that lead to systems like \J}\label{Sect:JlikeSystems}

We synthesize the Galactic population of binaries like \J\, by
randomly selecting $10^4$ triples that survived the supernova
explosion of the previous \S, and continue their evolution in AMUSE
(see \S\,\ref{Sect:NumericalMethod}).  Instead of performing a
self-consistent evolution as adopted in
\S\,\ref{Sect:SimulationResults} to validate the proposed scenario, we
tentatively decoupled the mass transfer process from the orbit
integration.  The population synthesis simulations start by resolving
the mass transfer in the inner binary until the triple becomes
dynamically unstable \citep{2001MNRAS.321..398M}, after which we
continue the simulation by resolving the dynamics of the 3-body system
until one star is ejected.  During this latter part we ignore the mass
transfer.

Note that this decoupled approach, though computationally cheaper by
about a factor of $10^3$, is not a-priori less reliable than the
self-consistent simulations in \S\,\ref{Sect:SimulationResults},
because our numerical methods provide no self consistent way to
resolve the mass transfer process in eccentric orbits. 

The majority ($\apgt 90\%$) of our simulations lead to the disruption of
the triple, which results in the ejection of the initial primary (MSP),
the secondary (partially stripped sub-giant or white dwarf) or the
tertiary (main-sequence F-, G- or K-dwarf) star, leaving the other two
stars in a binary.  The respective ratio at which these occur are
0.013, 0.683 and 0.304 for circular outer orbits (which is unlikely
after the supernova) to 0.03, 0.489, and 0.481 for highly eccentric
($e\sim 0.9$) orbits.  For extremely high eccentricities ($e \sim
0.99$) these fractions become 0.05, 0.75 and 0.20, respectively.  The
outer orbits tend to have high eccentricities because they often
barely survive the supernova kick, and as a consequence the total
fraction of systems in which the MSP remains in a binary with the
original outer component is $\aplt 0.5$; the fraction of single MSPs
is $\sim 0.05$.

The two methods through which we resolved the triples
(self-consistent, see \S\,\ref{Sect:NumericalMethod}, as opposed the
here adopted decoupled approach) give slightly different results, but
these can be related to the variation of the implementations. The
biggest effect is illustrated in Fig.\,\ref{Fig:ExampleEvolution},
where we demonstrated how the slow changes in the inner orbit drive
the secular evolution of the outer orbit. The main consequences of
this coupling are the higher mass of the neutron star by the time the
triple becomes dynamically unstable and the small fraction of triples
that survive the entire evolution because the adiabatic expansion of
the outer orbit prevents the triple from becoming dynamically
unstable. The consequential loss of systems however, is well
compensated by the increased probability that the donor in the LMXB is
ejected when the triple becomes dynamically unstable.  The differences
between the two numerical implementations affect our estimates for the
birthrate (see \S\,\ref{Sect:Birthrate}) on a $\aplt 20$\% level.  For
clarity, the statistical uncertainties between the two different
numerical approaches depend on Poissonian arguments rather than on the
lack of our understanding of parts of the physical process, such as
the common envelop evolution, the process of mass transfer in
non-circular orbits and the non-linear effects in the dynamically
unstable configuration just before the triple is resolved.  We control
the latter by adopting a high order and numerically extremely precise
algorithm (see \S\,\ref{Sect:NumericalMethod}). However, the
tipping-point physics of ejecting the MSP, its close white-dwarf
companion or the main-sequence outer star remains elusive, even when
integrating near machine precision.

\subsection{The number of \J-like systems in the Galaxy}\label{Sect:Birthrate}

The ratio of the birth rates of ordinary LMXBs to systems like \J\, is
a combination of the factors derived above: a few per cent for the
number of suitable triples, a reduction of a factor of $\sim 3$ owing
to the effect of an electron-capture supernova, and then a fraction
$\aplt 0.5$ of systems in which the inner secondary is ejected. For
larger average kick velocities
\citep{1998ApJ...505..315C,2002ApJ...568..289A} the fraction of
triples that survive the supernova drops from 1/3 to 1/50 (see
\S\,\ref{Sect:TripleSurvivors}).  The higher kicks cause an even
stronger reduction in the number of single MSPs, because outer orbits
tend to be highly eccentric, which leads to a reduction of the
probability that the MSP is ejected once the triple becomes
dynamically unstable.  The birth rate for systems like \J\, is then in
the range of $2 \times 10^{-4}$ to $3 \times 10^{-3}$ times the birth
rate of ordinary LMXBs. Estimates of the total number of LMXBs in the
Galaxy are in the range $10^4$ -- $10^5$
\citep{1989A&A...218..131C}. The typical life time of a LMXB is of the
order of a Gyr, whereas the life time of systems like \J\, is limited
by the life time of the MSP and the difference between that of the
inner secondary star and the outer tertiary star. Both stars have a
mass of the order of 1\,\Msun\, and their lifetimes exceed several
Gyr, which is a factor of a few longer than the lifetime of the LMXB
phase. The number of \J-like systems in the Galaxy then is at least
30--300 for electron-capture supernovae but drops to 3--30 when we
adopt considerably more mass to be lost in the supernova and higher
velocity kicks \citep{1998ApJ...505..315C,2002ApJ...568..289A}.

\section{Discussion}\label{Sect:Discussion}

We went through considerable effort to explain the existence of \J\,
as the result of the complex evolution in a hierarchical triple star
system.  In this section we will argue that though, unlikely as it
sounds, our proposed model is currently the only viable model
available, and that all other existing models fail to explain all the
characteristics of \J. After that we will indicate a number of other
sources that have evolved in quite a similar way, and that therefore
support the triple scenario.

\subsection{Shortcomings of earlier models for \J}\label{Sect:ModelsFailJ}

In the triple scenario for the formation of \J\, proposed by
\cite{2008Sci...320.1309C}, the observed $\sim 95$\,day orbital period
is that of an unobserved massive 0.9-1.1\,\Msun\, white dwarf, while
the observed G-dwarf has a considerably wider orbit. The observed
eccentricity of the inner orbit would in this case be driven by the
secular evolution via the Kozai mechanism \citep{1962AJ.....67..591K}.
This scenario has a number of serious shortcomings, one of which is
the unusually high white-dwarf mass, which leaves little room to
produce a massive 1.67\,\Msun\, neutron star. Since it's discovery,
the observed change in eccentricity ${\dot e} \sim 10^{-16} s^{-1}$ is
three orders of magnitude smaller than predictions based on the Kozai
mechanism \citep{2009MNRAS.399L.123G}, which excludes the presence of
a highly inclined tertiary star. We exclude this scenario therefore
from further consideration.

An alternative to the previous model would be the direct formation of
a MSP in a supernova explosion, or via the fall-back of material in a
circum neutron-star disk \citep{2009ApJ...692..723L}.  A variety of
arguments against the direct formation of rapidly spinning pulsar with
a small surface magnetic field ($2 \times 10^8$\,G) was provided by
\cite{2008Sci...320.1309C}. These include: first, of the 50 neutron
stars in young supernova remnants, none are fast-spinning low-field
pulsars \citep{2002ASPC..271....3K}. Second, a "born-fast" scenario for
\J\, would likely account for the 15 isolated MSPs detected in
the galactic disk, but the spin distribution, space velocities and
energetics of these single MSPs are indistinguishable from those of recycled, not "born
fast" binary MSPs \citep{2009Sci...324.1411A}; while their space
velocities and scale heights do not match those of non-recycled single
pulsars \citep{2007MNRAS.379..282L}. Third, magnetic fields in young
pulsars likely originate either from dynamo action in the proto-NS
\citep{1993ApJ...408..194T} or through compression of ``frozen-in''
fields of the progenitor star during collapse
\citep{2006MNRAS.367.1323F}, and no young pulsars with $B \aplt
10^{10}$\, Gauss, like \J, are known. We add a fourth argument: before
the direct collapse, the NS progenitor ascends the giant branch star
and common-envelope evolution with the 95\,day-orbit G-dwarf would
have dramatically reduced the orbital period, in the same way as
normal LMXBs form. For these four reasons, a core-collapse "born-fast" MSP
can be ruled out.

The above arguments hold equally well against the accretion-induced
collapse of a massive and rapidly rotating white dwarf to a neutron
star (NS). While this may lead to millisecond rotation periods, the NS
formed here will be just as hot and differentially rotating during the
early liquid phases as a NS formed by core collapse. In both cases
$10^{53}$\,erg in gravitational energy is released, which will erase
all memory of the violent mechanism in which the NS was
formed. Therefore, dynamo action in both cases will not depend on the
formation mechanism, and there is no fundamental reason for expecting
a weak magnetic field in NSs formed by accretion-induced collapse.
Furthermore, the direct accretion-induced collapse of a white dwarf to
a MSP in population studies produces binaries with a orbital period
$\aplt 20$\,days, which is considerably shorter than observed in \J\,
\citep{2010MNRAS.tmp.1408C}. In addition this scenario requires the
white dwarf to accrete from an evolved companion star, which is
inconsistent with the observed main-sequence companion G-star.  This
model thus fails to explain \J.

Finally, a formation scenario where \J\, is formed when the donor star
in the inner binary is ablated and destroyed, like is the case for the
``black-widow'' system PSR~B1957+20 \citep{1988Natur.333..237F}, has
thee problems. First, the observed timescales for straightforward
evaporation of the donor star are too long
\citep{2008Sci...320.1309C}. Second, formation of such a system likely
involves an exchange interaction \citep{2003MNRAS.345..678K}, which
would be greatly impeded by the outer G-dwarf companion. Third, even
if no exchange took place, the slow evaporation of the donor in the
inner binary requires the triple to be dynamically stable with respect
to the G-star in its current orbit.  Since the supernova explosion can
at most account for a reduction of a factor of 2 in the orbital
separation, the common-envelope should in that case be responsible for
a further reduction from the initial orbital separation of $\apgt
560$\,\Rsun\, to the currently observed $\sim 123$\,\Rsun, which
requires considerable fine tuning.  We therefore conclude that \J\, is
unlikely to have originated through a black-widow-like scenario.

Each of the above models has serious shortcomings, and we conclude
that none of the scenarios discussed above give a satisfactory
explanation for the formation of \J. 

\subsection{The missing link}\label{Sect:MissingLinks}

There are currently no triples known with parameters suitable for
evolving into systems like \J.  However, there is observational
evidence that such triples exist, as it is proposed that 4U~2129+47
(V1727 Cyg), a 5.24 hour LMXB, is accompanied by a spectral type
F-dwarf in an eccentric orbit of about 175\,days
\citep{1989ApJ...341L..75G,2008A&A...485..773B,2009ApJ...706.1069L}.
This triple can have formed in the same way as \J\, with the exception
that after the common-envelope and the subsequent supernova explosion
the inner binary period was smaller than the bifurcation period, of
about one day \citep{1988A&A...191...57P,1989A&A...208...52P}.  The
consequence of such a short orbital period is that the LMXB evolves to
an even shorter orbital period. Interestingly the 175\, day orbit of
the outer star is too small to have been dynamically stable at the
birth of the triple \citep{2001MNRAS.321..398M}. We argue that the
common-envelope and/or the supernova may have reduced the orbital
separation of the outer star.

We validated the probability that a binary orbit shrinks as a result
of the supernova by means of population synthesis and conclude that in
a fraction of 0.511 of the triples that survive the supernova the
separation of the circularized outer orbit is smaller than the initial
orbit.  However, if the progenitor of 4U~2129+47 had parameters
comparable to what we derived for \J\, in
\S\,\ref{Sect:ProposedModel}, the common-envelope phase of the inner
binary must have resulted in a reduction of the outer orbit as well,
by $\apgt 210$\,\Rsun.  Such a reduction in the separation of the
outer orbit by the common envelope enormously boosts the
surviveability of the triple in the supernova explosion (see
Fig.\,\ref{Fig:SupernovaSurvival}), and therefore dramatically
increases the birthrate of binaries like \J.

Following the same scenario as for forming 4U~2129+47 but with a less
massive ($\aplt 8$\,\Msun) initial inner primary star the inner binary
could evolve into a cataclysmic variable such as EC~19314-5915, with
an orbital period of 4.75\,hours. The observed radial velocity of
$\sim 9$\,km/s has been attributed to a G8-dwarf in orbit around the
CV \citep{1992MNRAS.258..285B}, which is consistent with a semi-major
axis of $\sim 2400$\,\Rsun\,

The real missing link would be the discovery of a relatively wide
($\apgt 30$\,\Rsun) LMXB that is orbited by a tertiary low-mass
main-sequence star. The mass-transfer phase in our simulations
averaged about 23\,Myr (see \S\,\ref{Sect:SimulationResults}), while
the lifetime of the MSP in \J\, is at least 1\,Gyr.  We therefore
expect that the Galaxy contains at most 7 such wide triple LMXBs.

\section{Conclusions}\label{Sect:Conclusions}

We discussed the evolution of triple star systems through a range of
dramatic events, including several tidal circularizations, a
common-envelope phase, a supernova and a stable phase of mass transfer
that eventually leads to a dynamical instability in which one star is
ejected.

In particular for producing a system like \J\, we require a triple to
be born as a rather ordinary dynamically stable hierarchical system of
which the inner binary consists of a 9--13\,\Msun\, and a
0.8--2.0\,\Msun\, star in a $\apgt 200$\,\Rsun\, separation. This
binary is orbited by a main sequence star with a mass smaller than the
initial secondary ($<2.0$\,\Msun) with a semi-major axis $\apgt
560$\,\Rsun.  The chance that the triple survives the inevitable chain
of events is not large but the result is profound and provides a
satisfactory explanation for a number of known systems in the Galaxy,
including \J, 4U~2129+47 (see \S\,\ref{Sect:MissingLinks}) and EC
19314-5915.

The range in possible observable stages in the evolutionary sequence
for forming a system like \J\, sensitively depends on the orbital
separation of the inner binary after the supernova, which makes the
distinction between evolving into a binary MSP like \J\, or a LMXB
with an outer tertiary companion like 4U~2129+47.  The evolution of
the triple naturally leads to a cataclysmic variable like
EC~19314-4915, if in addition to a short post-common envelope period
the initial primary star evolves into a white-dwarf rather than a
neutron star. We expect that such triple CV's are rather common in the
Galaxy.

We demonstrated that our model can indeed reproduce \J\,
qualitatively, and we estimate the number in the Galaxy by
performing extensive population synthesis of post-common envelope
triple systems (see \S\,\ref{Sect:HowManyJsInGalaxy}). Our starting
conditions are the progenitor of a LMXB which is accompanied by a third
low-mass companion in a relatively wide orbit. Based on the observed
statistics for such systems, their survival in the electron-capture
supernova explosion of an ONeMg star and the consequences of the
dynamical instability which results from the mass transfer in the
inner binary, we conclude that the formation rate of \J-like systems
is $\sim 3 \times 10^{-3}$ times that of LMXBs. With a life time at
least as long as that of LMXBs, and an estimated total number of
$10^4$ -- $10^5$ LMXBs in the Galaxy \citep{1989A&A...218..131C}, we
expect at least $30$--$300$ systems like \J\, in the Galaxy and an
order of magnitude smaller number of single MSPs. The longer lifetime
of the MSP binary compared to LMXB's results in an increase of this
number of a factor of a few. In the most pessimistic scenario, when we
adopt a higher velocity kick, this number drops to about 3--30 MSP
binaries like \J\, in the Galaxy, and a few single MSPs.

With a birthrate for Galactic LMXBs of $3.2 \times 10^{-6}$/yr
\citep{1998ApJ...493..351K} to $7 \times 10^{-6}$/yr
\citep{1989A&A...218..131C} we conclude that systems like \J\, form at
a rate of $\aplt 2.1 \times 10^{-8}$, and a ten times smaller rate for
single MSPs.  These low rates makes the proposed scenario unlikely, as
we already expected, but sufficiently probable that the Galaxy should
contain a few tens to hundreds of objects with characteristics similar
to \J\, and consequently provides a satisfactory explanation for \J.
Since the birthrate of single MSPs is expected to be quite similar to
that of LMXBs \citep{2010ScChG..53S.125D} our proposed triple scenario
does not significantly contribute to the formation of single MSPs.

\section*{Acknowledgments}
This work was supported by the Netherlands Research Council NWO (via
grants \#643.200.503, \#639.073.803 and \#614.061.608), the European
Commission (grant FP7-PEOPLE-2007-4-3-IRG \#224838) and the
Netherlands Research School for Astronomy (NOVA) for their support to
AMUSE.  Computing resources were provided by ASTRON.  We thank Ron
Taam for insightful discussions on common envelope evolution at the
MODEST-10 meeting in Beijing, and the AMUSE development team, Arjen
van Elteren, Inti Pelupessi, Nathan de Vries and Marcel Marosvolgyi,
for their much appreciated support, assistance and feedback.  We thank
the editor of Nature Leslie Sage and for anonymous Nature referees for
thoughtfull comments on an earlier version of this paper sumitted on
15 September 2010.



\begin{thebibliography}{67}
\expandafter\ifx\csname natexlab\endcsname\relax\def\natexlab#1{#1}\fi
\expandafter\ifx\csname url\endcsname\relax
  \def\url#1{\texttt{#1}}\fi
\expandafter\ifx\csname urlprefix\endcsname\relax\def\urlprefix{URL }\fi

\bibitem[{{Bhattacharya} \& {van den Heuvel}(1991)}]{1991PhR...203....1B}
{Bhattacharya}, D. \& {van den Heuvel}, E.~P.~J.
\newblock {Formation and evolution of binary and millisecond radio pulsars}.
\newblock \emph{\physrep} \textbf{203}, 1--124 (1991).

\bibitem[{{Lorimer}(2008)}]{LorimerLivingReview2008}
{Lorimer}, D.
\newblock Living Reviews.
\newblock In \emph{Relativity}, vol.~11, 8 (2008).

\bibitem[{{Cordes} \& {Chernoff}(1998)}]{1998ApJ...505..315C}
{Cordes}, J.~M. \& {Chernoff}, D.~F.
\newblock {Neutron Star Population Dynamics. II. Three-dimensional Space
  Velocities of Young Pulsars}.
\newblock \emph{\apj} \textbf{505}, 315--338 (1998).

\bibitem[{{Freire} \emph{et~al.}(2010)}]{2010arXiv1011.5809F}
{Freire}, P.~C.~C. \emph{et~al.}
\newblock {On the nature and evolution of the unique binary pulsar J1903+0327}.
\newblock \emph{ArXiv e-prints}  (2010).

\bibitem[{{Eggleton} \& {Kiseleva}(1996)}]{1996epbs.conf..345E}
{Eggleton}, P.~P. \& {Kiseleva}, L.~G.
\newblock {Stellar and Dynamical Evolution within Triple Stars}.
\newblock In \emph{NATO ASIC Proc. 477: Evolutionary Processes in Binary Stars}
  (ed. {R.~A.~M.~J.~Wijers, M.~B.~Davies, \& C.~A.~Tout}), 345--+ (1996).

\bibitem[{{Eggleton}(2006)}]{2006epbm.book.....E}
{Eggleton}, P.
\newblock \emph{{Evolutionary Processes in Binary and Multiple Stars}}
  (Evolutionary Processes in Binary and Multiple Stars, by Peter Eggleton,
  pp.~.~ISBN 0521855578.~Cambridge, UK: Cambridge University Press, 2006.,
  2006).

\bibitem[{{Webbink}(1984)}]{1984ApJ...277..355W}
{Webbink}, R.~F.
\newblock Double white dwarfs as progenitors of R Coronae Borealis stars and
  Type I supernovae.
\newblock \emph{\apj} \textbf{277}, 355--360 (1984).

\bibitem[{{Taam} \& {Sandquist}(2000)}]{2000ARA&A..38..113T}
{Taam}, R.~E. \& {Sandquist}, E.~L.
\newblock {Common Envelope Evolution of Massive Binary Stars}.
\newblock \emph{\araa} \textbf{38}, 113--141 (2000).

\bibitem[{{Deloye} \& {Taam}(2010)}]{2010ApJ...719L..28D}
{Deloye}, C.~J. \& {Taam}, R.~E.
\newblock {Adiabatic Mass Loss and the Outcome of the Common Envelope Phase of
  Binary Evolution}.
\newblock \emph{\apjl} \textbf{719}, L28--L31 (2010).

\bibitem[{{Ge} \emph{et~al.}(2010){Ge}, {Hjellming}, {Webbink}, {Chen} \&
  {Han}}]{2010ApJ...717..724G}
{Ge}, H., {Hjellming}, M.~S., {Webbink}, R.~F., {Chen}, X. \& {Han}, Z.
\newblock {Adiabatic Mass Loss in Binary Stars. I. Computational Method}.
\newblock \emph{\apj} \textbf{717}, 724--738 (2010).

\bibitem[{{Blaauw}(1961)}]{1961BAN....15..265B}
{Blaauw}, A.
\newblock {On the origin of the O- and B-type stars with high velocities (the
  "run-away" stars), and some related problems}.
\newblock \emph{\bain} \textbf{15}, 265--+ (1961).

\bibitem[{{Dewey} \& {Cordes}(1987)}]{1987ApJ...321..780D}
{Dewey}, R.~J. \& {Cordes}, J.~M.
\newblock {Monte Carlo simulations of radio pulsars and their progenitors}.
\newblock \emph{\apj} \textbf{321}, 780--798 (1987).

\bibitem[{{Dewi} \emph{et~al.}(2005){Dewi}, {Podsiadlowski} \&
  {Pols}}]{2005MNRAS.363L..71D}
{Dewi}, J.~D.~M., {Podsiadlowski}, P. \& {Pols}, O.~R.
\newblock {The spin period-eccentricity relation of double neutron stars:
  evidence for weak supernova kicks?}
\newblock \emph{\mnras} \textbf{363}, L71--L75 (2005).

\bibitem[{{Webbink} \emph{et~al.}(1983){Webbink}, {Rappaport} \&
  {Savonije}}]{1983ApJ...270..678W}
{Webbink}, R.~F., {Rappaport}, S. \& {Savonije}, G.~J.
\newblock {On the evolutionary status of bright, low-mass X-ray sources}.
\newblock \emph{\apj} \textbf{270}, 678--693 (1983).

\bibitem[{{Taam}(1983)}]{1983ApJ...270..694T}
{Taam}, R.~E.
\newblock {The evolution of a stripped giant-neutron star binary}.
\newblock \emph{\apj} \textbf{270}, 694--699 (1983).

\bibitem[{{Garcia} \emph{et~al.}(1989){Garcia}, {Bailyn}, {Grindlay} \&
  {Molnar}}]{1989ApJ...341L..75G}
{Garcia}, M.~R., {Bailyn}, C.~D., {Grindlay}, J.~E. \& {Molnar}, L.~A.
\newblock {Is 4U 2129 + 47 (= V1727 Cygni) a triple system?}
\newblock \emph{\apjl} \textbf{341}, L75--L78 (1989).

\bibitem[{{Bothwell} \emph{et~al.}(2008){Bothwell}, {Torres}, {Garcia} \&
  {Charles}}]{2008A&A...485..773B}
{Bothwell}, M.~S., {Torres}, M.~A.~P., {Garcia}, M.~R. \& {Charles}, P.~A.
\newblock {Spectroscopic observations of the quiescent neutron star system 4U
  2129+47 (=V1727 Cygni)}.
\newblock \emph{\aap} \textbf{485}, 773--775 (2008).

\bibitem[{{Lin} \emph{et~al.}(2009){Lin}, {Nowak} \&
  {Chakrabarty}}]{2009ApJ...706.1069L}
{Lin}, J., {Nowak}, M.~A. \& {Chakrabarty}, D.
\newblock {A Further Drop into Quiescence by the Eclipsing Neutron Star 4U
  2129+47}.
\newblock \emph{\apj} \textbf{706}, 1069--1077 (2009).

\bibitem[{{Pylyser} \& {Savonije}(1988)}]{1988A&A...191...57P}
{Pylyser}, E. \& {Savonije}, G.~J.
\newblock {Evolution of low-mass close binary sytems with a compact mass
  accreting component}.
\newblock \emph{\aap} \textbf{191}, 57--70 (1988).

\bibitem[{{Pylyser} \& {Savonije}(1989)}]{1989A&A...208...52P}
{Pylyser}, E.~H.~P. \& {Savonije}, G.~J.
\newblock {The evolution of low-mass close binary systems with a compact
  component. II - Systems captured by angular momentum losses}.
\newblock \emph{\aap} \textbf{208}, 52--62 (1989).

\bibitem[{{Alpar} \emph{et~al.}(1982){Alpar}, {Cheng}, {Ruderman} \&
  {Shaham}}]{1982Natur.300..728A}
{Alpar}, M.~A., {Cheng}, A.~F., {Ruderman}, M.~A. \& {Shaham}, J.
\newblock {A new class of radio pulsars}.
\newblock \emph{\nat} \textbf{300}, 728--730 (1982).

\bibitem[{{Radhakrishnan} \& {Srinivasan}(1982)}]{1982CSci...51.1096R}
{Radhakrishnan}, V. \& {Srinivasan}, G.
\newblock {On the origin of the recently discovered ultra-rapid pulsar}.
\newblock \emph{Current Science} \textbf{51}, 1096--1099 (1982).

\bibitem[{{Heggie}(1975)}]{1975MNRAS.173..729H}
{Heggie}, D.~C.
\newblock Binary evolution in stellar dynamics.
\newblock \emph{\mnras} \textbf{173}, 729--787 (1975).

\bibitem[{{Fregeau} \emph{et~al.}(2004){Fregeau}, {Cheung}, {Portegies Zwart}
  \& {Rasio}}]{2004MNRAS.352....1F}
{Fregeau}, J.~M., {Cheung}, P., {Portegies Zwart}, S.~F. \& {Rasio}, F.~A.
\newblock {Stellar collisions during binary-binary and binary-single star
  interactions}.
\newblock \emph{\mnras} \textbf{352}, 1--19 (2004).

\bibitem[{{Mardling} \& {Aarseth}(2001)}]{2001MNRAS.321..398M}
{Mardling}, R.~A. \& {Aarseth}, S.~J.
\newblock {Tidal interactions in star cluster simulations}.
\newblock \emph{\mnras} \textbf{321}, 398--420 (2001).

\bibitem[{{Schwab} \emph{et~al.}(2010){Schwab}, {Podsiadlowski} \&
  {Rappaport}}]{2010ApJ...719..722S}
{Schwab}, J., {Podsiadlowski}, P. \& {Rappaport}, S.
\newblock {Further Evidence for the Bimodal Distribution of Neutron-star
  Masses}.
\newblock \emph{\apj} \textbf{719}, 722--727 (2010).

\bibitem[{{Shore} \emph{et~al.}(1994){Shore}, {Livio} \& {van den
  Heuvel}}]{1994inbi.conf.....S}
{Shore}, S.~N., {Livio}, M. \& {van den Heuvel}, E.~P.~J.
\newblock {Interacting binaries}.
\newblock In \emph{Saas-Fee Advanced Course 22: Interacting Binaries} (ed.
  {S.~N.~Shore, M.~Livio, \& E.~P.~J.~van den Heuvel}) (1994).

\bibitem[{{Portegies Zwart}(1995)}]{1995A&A...296..691P}
{Portegies Zwart}, S.~F.
\newblock {The formation of Be stars in close binary systems. The importance of
  kicks and angular-momentum loss}.
\newblock \emph{\aap} \textbf{296}, 691--+ (1995).

\bibitem[{{Portegies Zwart} \emph{et~al.}(2009)}]{2009NewA...14..369P}
{Portegies Zwart}, S. \emph{et~al.}
\newblock {A multiphysics and multiscale software environment for modeling
  astrophysical systems}.
\newblock \emph{New Astronomy} \textbf{14}, 369--378 (2009).

\bibitem[{{Aarseth} \& {Zare}(1974{\natexlab{a}})}]{1974CeMec..10..185A}
{Aarseth}, S.~J. \& {Zare}, K.
\newblock {A regularization of the three-body problem}.
\newblock \emph{Celestial Mechanics} \textbf{10}, 185--205
  (1974{\natexlab{a}}).

\bibitem[{{Aarseth} \& {Zare}(1974{\natexlab{b}})}]{1974CeMec..10..516A}
{Aarseth}, S.~J. \& {Zare}, K.
\newblock {Errata: ''A regularization of the three-body problem'' [Celestial
  Mech., Vol. 10, p. 185 - 205 (1974)].}
\newblock \emph{Celestial Mechanics} \textbf{10}, 516--+ (1974{\natexlab{b}}).

\bibitem[{{Tauris} \& {Savonije}(1999)}]{1999A&A...350..928T}
{Tauris}, T.~M. \& {Savonije}, G.~J.
\newblock {Formation of millisecond pulsars. I. Evolution of low-mass X-ray
  binaries with $P_{orb} > 2$ days}.
\newblock \emph{\aap} \textbf{350}, 928--944 (1999).

\bibitem[{{Kozai}(1979)}]{1979IAUS...81..231K}
{Kozai}, Y.
\newblock {Secular perturbations of asteroids and comets}.
\newblock In \emph{IAU Symp. 81: Dynamics of the Solar System} (ed. {Duncombe},
  R.~L.), 231--236 (1979).

\bibitem[{{Sepinsky} \emph{et~al.}(2009){Sepinsky}, {Willems}, {Kalogera} \&
  {Rasio}}]{2009ApJ...702.1387S}
{Sepinsky}, J.~F., {Willems}, B., {Kalogera}, V. \& {Rasio}, F.~A.
\newblock {Interacting Binaries with Eccentric Orbits. II. Secular Orbital
  Evolution due to Non-conservative Mass Transfer}.
\newblock \emph{\apj} \textbf{702}, 1387--1392 (2009).

\bibitem[{{Sepinsky} \emph{et~al.}(2010){Sepinsky}, {Willems}, {Kalogera} \&
  {Rasio}}]{2010ApJ...724..546S}
{Sepinsky}, J.~F., {Willems}, B., {Kalogera}, V. \& {Rasio}, F.~A.
\newblock {Interacting Binaries with Eccentric Orbits. III. Orbital Evolution
  due to Direct Impact and Self-Accretion}.
\newblock \emph{\apj} \textbf{724}, 546--558 (2010).

\bibitem[{{Lajoie} \& {Sills}(2010{\natexlab{a}})}]{2010arXiv1011.2204L}
{Lajoie}, C. \& {Sills}, A.
\newblock {Mass Transfer in Binary Stars using SPH. II. Eccentric Binaries}.
\newblock \emph{ArXiv e-prints}  (2010{\natexlab{a}}).

\bibitem[{{Lajoie} \& {Sills}(2010{\natexlab{b}})}]{2010arXiv1011.2211L}
{Lajoie}, C. \& {Sills}, A.
\newblock {Mass Transfer in Binary Stars using SPH. I. Numerical Method}.
\newblock \emph{ArXiv e-prints}  (2010{\natexlab{b}}).

\bibitem[{{Eggleton} \& {Tokovinin}(2008)}]{2008MNRAS.389..869E}
{Eggleton}, P.~P. \& {Tokovinin}, A.~A.
\newblock {A catalogue of multiplicity among bright stellar systems}.
\newblock \emph{\mnras} \textbf{389}, 869--879 (2008).

\bibitem[{{Eggleton} \& {Tokovinin}(2010)}]{2010yCat..73890869E}
{Eggleton}, P.~P. \& {Tokovinin}, A.~A.
\newblock {Multiplicity among bright stellar systems (Eggleton+, 2008)}.
\newblock \emph{VizieR Online Data Catalog} \textbf{738}, 90869--+ (2010).

\bibitem[{{Boersma}(1961)}]{1961BAN....15..291B}
{Boersma}, J.
\newblock {Mathematical theory of the two-body problem with one of the masses
  decreasing with time}.
\newblock \emph{\bain} \textbf{15}, 291--301 (1961).

\bibitem[{{Hills}(1983)}]{1983ApJ...267..322H}
{Hills}, J.~G.
\newblock {The effects of sudden mass loss and a random kick velocity produced
  in a supernova explosion on the dynamics of a binary star of arbitrary
  orbital eccentricity - Applications to X-ray binaries and to the binary
  pulsars}.
\newblock \emph{\apj} \textbf{267}, 322--333 (1983).

\bibitem[{{Brandt} \& {Podsiadlowski}(1995)}]{1995MNRAS.274..461B}
{Brandt}, N. \& {Podsiadlowski}, P.
\newblock {The effects of high-velocity supernova kicks on the orbital
  properties and sky distributions of neutron-star binaries}.
\newblock \emph{\mnras} \textbf{274}, 461--484 (1995).

\bibitem[{{Tauris} \& {Takens}(1998)}]{1998A&A...330.1047T}
{Tauris}, T.~M. \& {Takens}, R.~J.
\newblock {Runaway velocities of stellar components originating from disrupted
  binaries via asymmetric supernova explosions}.
\newblock \emph{\aap} \textbf{330}, 1047--1059 (1998).

\bibitem[{{Podsiadlowski} \emph{et~al.}(2004)}]{2004ApJ...612.1044P}
{Podsiadlowski}, P. \emph{et~al.}
\newblock {The Effects of Binary Evolution on the Dynamics of Core Collapse and
  Neutron Star Kicks}.
\newblock \emph{\apj} \textbf{612}, 1044--1051 (2004).

\bibitem[{{Scheck} \emph{et~al.}(2004){Scheck}, {Plewa}, {Janka}, {Kifonidis}
  \& {M{\"u}ller}}]{2004PhRvL..92a1103S}
{Scheck}, L., {Plewa}, T., {Janka}, H., {Kifonidis}, K. \& {M{\"u}ller}, E.
\newblock {Pulsar Recoil by Large-Scale Anisotropies in Supernova Explosions}.
\newblock \emph{Physical Review Letters} \textbf{92}, 011103--+ (2004).

\bibitem[{{Lyne} \& {Lorimer}(1994)}]{1994Natur.369..127L}
{Lyne}, A.~G. \& {Lorimer}, D.~R.
\newblock High Birth Velocities of Radio Pulsars.
\newblock \emph{\nat} \textbf{369}, 127+ (1994).

\bibitem[{{Hansen} \& {Phinney}(1997)}]{1997MNRAS.291..569H}
{Hansen}, B.~M.~S. \& {Phinney}, E.~S.
\newblock {The pulsar kick velocity distribution}.
\newblock \emph{\mnras} \textbf{291}, 569--+ (1997).

\bibitem[{{Arzoumanian} \emph{et~al.}(2002){Arzoumanian}, {Chernoff} \&
  {Cordes}}]{2002ApJ...568..289A}
{Arzoumanian}, Z., {Chernoff}, D.~F. \& {Cordes}, J.~M.
\newblock {The Velocity Distribution of Isolated Radio Pulsars}.
\newblock \emph{\apj} \textbf{568}, 289--301 (2002).

\bibitem[{{van den Heuvel}(2004)}]{2004ESASP.552..185V}
{van den Heuvel}, E.~P.~J.
\newblock {X-Ray Binaries and Their Descendants: Binary Radio Pulsars; Evidence
  for Three Classes of Neutron Stars?}
\newblock In \emph{5th INTEGRAL Workshop on the INTEGRAL Universe} (ed.
  {V.~Schoenfelder, G.~Lichti, \& C.~Winkler}), vol. 552 of \emph{ESA Special
  Publication}, 185--+ (2004).

\bibitem[{{Hobbs} \emph{et~al.}(2005){Hobbs}, {Lorimer}, {Lyne} \&
  {Kramer}}]{2005MNRAS.360..974H}
{Hobbs}, G., {Lorimer}, D.~R., {Lyne}, A.~G. \& {Kramer}, M.
\newblock {A statistical study of 233 pulsar proper motions}.
\newblock \emph{\mnras} \textbf{360}, 974--992 (2005).

\bibitem[{{Willems} \& {Kolb}(2003)}]{2003MNRAS.343..949W}
{Willems}, B. \& {Kolb}, U.
\newblock {On the detection of pre-low-mass X-ray binaries}.
\newblock \emph{\mnras} \textbf{343}, 949--958 (2003).

\bibitem[{{Cote} \& {Pylyser}(1989)}]{1989A&A...218..131C}
{Cote}, J. \& {Pylyser}, E.~H.~P.
\newblock {The birthrates of galactic low mass binary radio pulsars and their
  progenitor systems}.
\newblock \emph{\aap} \textbf{218}, 131--136 (1989).

\bibitem[{{Champion} \emph{et~al.}(2008)}]{2008Sci...320.1309C}
{Champion}, D.~J. \emph{et~al.}
\newblock {An Eccentric Binary Millisecond Pulsar in the Galactic Plane}.
\newblock \emph{Science} \textbf{320}, 1309-- (2008).

\bibitem[{{Kozai}(1962)}]{1962AJ.....67..591K}
{Kozai}, Y.
\newblock {Secular perturbations of asteroids with high inclination and
  eccentricity}.
\newblock \emph{\aj} \textbf{67}, 591--+ (1962).

\bibitem[{{Gopakumar} \emph{et~al.}(2009){Gopakumar}, {Bagchi} \&
  {Ray}}]{2009MNRAS.399L.123G}
{Gopakumar}, A., {Bagchi}, M. \& {Ray}, A.
\newblock {Ruling out Kozai resonance in highly eccentric galactic binary
  millisecond pulsar PSR J1903+0327}.
\newblock \emph{\mnras} \textbf{399}, L123--L127 (2009).

\bibitem[{{Liu} \& {Li}(2009)}]{2009ApJ...692..723L}
{Liu}, X. \& {Li}, X.
\newblock {A Fallback Disk Accretion Involved Formation Channel to PSR
  J1903+0327}.
\newblock \emph{\apj} \textbf{692}, 723--728 (2009).

\bibitem[{{Kaspi} \& {Helfand}(2002)}]{2002ASPC..271....3K}
{Kaspi}, V.~M. \& {Helfand}, D.~J.
\newblock {Constraining the Birth Events of Neutron Stars}.
\newblock In \emph{Neutron Stars in Supernova Remnants} (ed. {P.~O.~Slane \&
  B.~M.~Gaensler}), vol. 271 of \emph{Astronomical Society of the Pacific
  Conference Series}, 3--+ (2002).

\bibitem[{{Archibald} \emph{et~al.}(2009)}]{2009Sci...324.1411A}
{Archibald}, A.~M. \emph{et~al.}
\newblock {A Radio Pulsar/X-ray Binary Link}.
\newblock \emph{Science} \textbf{324}, 1411-- (2009).

\bibitem[{{Lorimer} \emph{et~al.}(2007){Lorimer}, {McLaughlin}, {Champion} \&
  {Stairs}}]{2007MNRAS.379..282L}
{Lorimer}, D.~R., {McLaughlin}, M.~A., {Champion}, D.~J. \& {Stairs}, I.~H.
\newblock {PSR J1453+1902 and the radio luminosities of solitary versus binary
  millisecond pulsars}.
\newblock \emph{\mnras} \textbf{379}, 282--288 (2007).

\bibitem[{{Thompson} \& {Duncan}(1993)}]{1993ApJ...408..194T}
{Thompson}, C. \& {Duncan}, R.~C.
\newblock {Neutron star dynamos and the origins of pulsar magnetism}.
\newblock \emph{\apj} \textbf{408}, 194--217 (1993).

\bibitem[{{Ferrario} \& {Wickramasinghe}(2006)}]{2006MNRAS.367.1323F}
{Ferrario}, L. \& {Wickramasinghe}, D.
\newblock {Modelling of isolated radio pulsars and magnetars on the fossil
  field hypothesis}.
\newblock \emph{\mnras} \textbf{367}, 1323--1328 (2006).

\bibitem[{{Chen} \emph{et~al.}(2010){Chen}, {Liu}, {Xu} \&
  {Li}}]{2010MNRAS.tmp.1408C}
{Chen}, W., {Liu}, X., {Xu}, R. \& {Li}, X.
\newblock {Can eccentric binary millisecond pulsars form by accretion-induced
  collapse of white dwarfs?}
\newblock \emph{\mnras} 1408--+ (2010).

\bibitem[{{Fruchter} \emph{et~al.}(1988){Fruchter}, {Stinebring} \&
  {Taylor}}]{1988Natur.333..237F}
{Fruchter}, A.~S., {Stinebring}, D.~R. \& {Taylor}, J.~H.
\newblock {A millisecond pulsar in an eclipsing binary}.
\newblock \emph{\nat} \textbf{333}, 237--239 (1988).

\bibitem[{{King} \emph{et~al.}(2003){King}, {Davies} \&
  {Beer}}]{2003MNRAS.345..678K}
{King}, A.~R., {Davies}, M.~B. \& {Beer}, M.~E.
\newblock {Black widow pulsars: the price of promiscuity}.
\newblock \emph{\mnras} \textbf{345}, 678--682 (2003).

\bibitem[{{Buckley} \emph{et~al.}(1992){Buckley}, {O'Donoghue}, {Kilkenny},
  {Stobie} \& {Remillard}}]{1992MNRAS.258..285B}
{Buckley}, D.~A.~H., {O'Donoghue}, D., {Kilkenny}, D., {Stobie}, R.~S. \&
  {Remillard}, R.~A.
\newblock {EC 19314 - 5915 - A bright, eclipsing cataclysmic variable from the
  Edinburgh-Cape Blue Object Survey}.
\newblock \emph{\mnras} \textbf{258}, 285--295 (1992).

\bibitem[{{Kalogera} \& {Webbink}(1998)}]{1998ApJ...493..351K}
{Kalogera}, V. \& {Webbink}, R.~F.
\newblock {Formation of Low-Mass X-Ray Binaries. II. Common Envelope Evolution
  of Primordial Binaries with Extreme Mass Ratios}.
\newblock \emph{\apj} \textbf{493}, 351--+ (1998).

\bibitem[{{Dai} \& {Li}(2010)}]{2010ScChG..53S.125D}
{Dai}, H. \& {Li}, X.
\newblock {The low-mass X-ray binary-millisecond radio pulsar birthrate problem
  revisited}.
\newblock \emph{Science in China G: Physics and Astronomy} \textbf{53},
  125--129 (2010).

\end{thebibliography}
\end{document}